\documentclass[11pt]{article}
\usepackage{amsfonts,amsmath,amssymb,epsf}
\usepackage{graphicx,color}
\newcommand{\al}{\alpha'}
\newcommand{\de}{\partial}
\newcommand{\be}{\begin{equation}}
\newcommand{\ba}{\begin{eqnarray}}
\newcommand{\ea}{\end{eqnarray}}
\newcommand{\ee}{\end{equation}}

\newcommand{\we}{\wedge}

\newcommand{\f}{\frac}
\newcommand{\s}{\sqrt}
\newcommand{\vp}{\varphi}

\newcommand{\ti}{\tilde}
\newcommand{\ap}{\alpha}

\newcommand{\ddd}{\cdot\cdot\cdot}
\newcommand{\no}{\nonumber \\}
\newcommand{\la}{\langle}
\newcommand{\lb}{\rangle}
\newcommand{\ep}{\epsilon}

\def\CR{{\cal R}}

\newcommand{\Slash}[1]{\ooalign{\hfil/\hfil\crcr$#1$}}

\begin{document}

\begin{titlepage}
\thispagestyle{empty}

KUNS-2147
\begin{flushright}
\end{flushright}

\bigskip

\begin{center}
\noindent{\Large \textbf{On Type IIA Penrose Limit and ${\cal N}=6$
Chern-Simons Theories
}}\\
\vspace{2cm} {Tatsuma
Nishioka\footnote{e-mail:nishioka@gauge.scphys.kyoto-u.ac.jp}
 and Tadashi
Takayanagi\footnote{e-mail:takayana@gauge.scphys.kyoto-u.ac.jp}}\\
\vspace{1cm}
 {\it  Department of Physics, Kyoto University, Kyoto 606-8502, Japan
 }

\vskip 3em
\end{center}

\begin{abstract}
Recently, Aharony, Bergman, Jafferis and Maldacena proposed that the
${\cal N}=6$ Chern-Simons gauge theories are holographically dual to
the M-theory backgrounds with multiple M2-branes on orbifolds
$C^4/Z_k$. When $k$ is large, they have the type IIA string
description. In this paper we analyze the Penrose limit of this IIA
background and express the string spectrum as the conformal
dimensions of operators in the gauge theories. For BPS operators, we
can confirm the agreements between the IIA string on plane waves and
the gauge theories. We point out that there exist BMN-like operators
in the gauge theories, though their holographic interpretation does
not seem to be simple. Also we analyze the weak coupling limit of
this theory and show that the Hagedorn/deconfinement transition
occurs as expected.

\end{abstract}

\end{titlepage}

\newpage
\section{Introduction}
\setcounter{equation}{0}
Recently, there have been very interesting progresses on the
world-volume theory of multiple M2-branes in M-theory. Since we
expect that this theory is dual to $AdS_4\times S^7$ in the
decoupling limit, it should be described by a three dimensional
${\cal N}=8$ superconformal theory. The superconformal Chern-Simons
theories have been constructed in \cite{Sc,GaYi,GaWi}, though they
are less supersymmetric. Bagger, Lambert and Gustavsson constructed
the three dimensional ${\cal N}=8$ supersymmetric theory based on
the Lie 3-algebra structure \cite{BaLa,Gu}. For subsequent
developments refer to \cite{MuA}-\cite{Blau}. If we assume the
positive metric of the Lie 3-algebra, the algebra constraints the
Lagrangian strongly \cite{LiA,LiB,LiC,LiD}. This only allows us to
construct a ${\cal N}=8$ supersymmetric theory which is dual to two
M2-branes.  If we allow the non-degenerate metric, we can find
${\cal N}=8$ supersymmetric theories where we can take the number of
branes $N$ arbitrary large \cite{DEA,DEB,DEC,DED,DEE,DEF,DEG}.
However, this theory can be reduced to the well-known ${\cal N}=8$
super Yang-Mills theory on $N$ $D2$-branes.

Aharony, Bergman, Jafferis and Maldacena have constructed very
interesting ${\cal N}=6$ $U(N)\times U(N)$ Chern-Simons theories
(ABJM theory) and proposed that they are dual to the world-volume
theory of $N$ M2-branes for the arbitrary number of $N$ \cite{ABJM}
(see also \cite{Kl,Mi} for further study). This theory is
parameterized by the level $k$ of the Chern-Simons gauge theory. The
ABJM theory with level $k$ and the gauge group $U(N)\times U(N)$ is
argued to describe the world-volume theory of M2-branes on the
orbifold $C^4/Z_k$. When the gauge group is $SU(2)\times SU(2)$,
this theory becomes equivalent to the BLG theory \cite{BaLa,Gu}.

In the paper \cite{ABJM}, it was also pointed out that the ABJM
theory at large $k$ is dual to the type IIA string on $AdS_4\times
CP^3$. This offers us to study a new $AdS_4/CFT_3$ duality where
both $AdS$ and $CFT$ side are tractable with the present knowledge
of string theory. The CFT side is defined by a 't Hooft limit
$N\to\infty$ of ABJM theory with $\f{N}{k}$ kept finite.

As a next step, it will be very intriguing to check this proposed
duality from both the IIA string theory and the Chern-Simons gauge
theory side. In this paper we would like to report a modest progress
in this direction. Namely, we would like to consider the Penrose
limit of the type IIA background because in this limit the string
theory becomes solvable even in the presence of $\al$ corrections
and RR-fluxes. It has been well-known that the Penrose limit of type
IIB string on $AdS_5\times S^5$ successfully reproduces the results
of BMN operators in the ${\cal N}=4$ super Yang-Mills theory
\cite{BMN}.

We will show that the Penrose limit of this IIA string background
becomes the plane wave background with 24 supersymmetries studied in
\cite{pp} after an appropriate coordinate transformation. We
find\footnote{The same plane-wave also appears in the study of the
gravity dual of a 2+1 super Yang-Mills with $SU(2|4)$ symmetry
\cite{LiMa}, where the string spectrum is compared with the
Yang-Mills operators.} the exact string spectrum and express the
results as the anomalous dimensions of operators in the ABJM theory.
We will also notice that in the ABJM theory, we can define a
BMN-like operator and we will compute its anomalous dimension to
leading order of the effective 't Hooft coupling.

We will also study the weak coupling limit $k\to \infty$ of the ABJM
theory. Since we can neglect the non-singlet flux contributions in
this limit, we can analyze the partition function of the ABJM theory
compactified on $S^1\times S^2$ analytically and show that the
deconfinement/confinement transition occurs at a specific
temperature as expected from the Hagedorn transition in the string
theory side.

This paper is organized as follows. In section 2, we review and
analyze in the detail the reduction of $AdS_4\times S^7$ background
in M-theory to the type IIA string. We will also compute the
holographic entanglement entropy of the ABJM theory. In section 3,
we take the Penrose limit of the IIA background. We compute the
string spectrum and express the results from the gauge theoretic
viewpoint. In section 4, we define BMN-like operators in ABJM theory
and compute their anomalous dimensions. In section 5, we evaluate
the partition of free ABJM theory and confirm the Hagedorn
transition at a specific temperature. In section 6 we summarize our
conclusions.

After we finished this paper, we noticed a very interesting preprint
\cite{Mi} by Bhattacharya and Minwalla, where an agreement of the
supersymmetric index between the ${\cal N}=6$ Chern-Simons theory
and its dual supergravity was shown. The section 5 of our paper has
 some overlap with their calculations.

\section{M2-branes on $C^4/Z_k$ and Reduction to IIA String}
\setcounter{equation}{0}
\subsection{M2-brane Solution}

We start with the 11 dimensional supergravity action \be
S=\f{1}{2\kappa_{11}^2}\int dx^{11}\s{-g}\left(R-\f{1}{2\cdot
4!}F_{\mu\nu\rho\sigma}F^{\mu\nu\rho\sigma}\right)-\f{1}{12\kappa_{11}^2}
\int C^{(3)}\we F^{(4)}\we F^{(4)}, \ee where
$\kappa_{11}^2=2^7\pi^8 l_p^9$. The equations of motions read \be
R^\mu_\nu=\f{1}{2}\left(\f{1}{3!}F^{\mu\ap\beta\gamma}F_{\nu\ap\beta\gamma}
-\f{1}{3\cdot 4!}\delta^\mu_\nu
F_{\ap\beta\rho\sigma}F^{\ap\beta\rho\sigma}\right), \ee and \be
\de_{\sigma}(\s{-g}F^{\sigma\mu\nu\xi})=\f{1}{2\cdot
(4!)^2}\ep^{\mu\nu\xi\ap_1\ddd\ap_8}F_{\ap_1\ddd\ap_4}F_{\ap_5\ddd\ap_8}.\ee
Then the near horizon limit of M2-brane solution becomes
$AdS_4\times S^7$ \be ds^2=\f{R^2}{4}ds^2_{AdS_4}+R^2 d\Omega_7^2,
\ee where the radius $R$ is given by $R=l_p(2^5 N'\pi^2)^{\f{1}{6}}$
($N'$ is the number of the M2-branes). The four form flux is found
to be \be F^{(4)}=\f{3R^3}{8}\ep_{AdS_4}, \ee where $\ep_{AdS_4}$ is
the unit volume form of the $AdS_4$ space. If we assume the Poincare
metric $ds^2_{AdS_4}=\f{dr^2}{r^2}+r^2\sum_{\mu=0}^2 dx^\mu dx_\mu$,
we have $\ep_{AdS_4}=r^2$ or equally $F_{012r}=\f{3R^3r^2}{8}$.

\subsection{The Reduction to IIA}

We take the $Z_k$ orbifold of $S^7$ and reduce the M-theory
background $AdS_4\times S^7/Z_{k}$ to the type IIA string background
following \cite{ABJM}. We can express $S^7$  by the complex
coordinate $X_1,X_2,X_3$ and $X_4$ with the constraint
$|X_1|^2+|X_2|^2+|X_3|^2+|X_4|^2=1$. We can parameterize $S^7$ by
\ba X_1&=&\cos\xi~ \cos\f{\theta_1}{2}~e^{i\f{\chi_1+\vp_1}{2}}, \no
X_2&=&\cos\xi~ \sin\f{\theta_1}{2}~e^{i\f{\chi_1-\vp_1}{2}}, \no
X_3&=&\sin\xi~ \cos\f{\theta_2}{2}~e^{i\f{\chi_2+\vp_2}{2}}, \no
X_4&=&\sin\xi~ \sin\f{\theta_2}{2}~e^{i\f{\chi_2-\vp_2}{2}},
\label{angles} \ea where the angular valuables run the values $0\leq
\xi <\f{\pi}{2}$, $0\leq \chi_i <4\pi$, $0\leq \vp_i \leq 2\pi$ and
$0\leq \theta_i<\pi$. Then the metric of $S^7$ can be written as \ba
ds_{S^7}^2&=&d\xi^2+\f{\cos^2\xi}{4}\left[(d\chi_1+\cos\theta_1d\vp_1)^2+
d\theta_1^2+\sin^2\theta_1 d\vp_1^2\right]\no & &\ \ \ +
\f{\sin^2\xi}{4}\left[(d\chi_2+\cos\theta_2 d\vp_2)^2+
d\theta_2^2+\sin^2\theta_2 d\vp_2^2\right]. \ea Now we define new
coordinates \be \chi_1=2y+\psi,\ \ \ \ \ \ \ \chi_2=2y-\psi. \ee The
$Z_k$ orbifold action is now given by $y\sim y+\f{2\pi}{k}$. Then
the metric of $S^7$ can be rewritten as follows \be
ds_{S^7}^2=ds^2_{CP^3}+(dy+A)^2, \ee where \ba
A&=&\f{1}{2}(\cos^2\xi-\sin^2\xi)d\psi +\f{1}{2}\cos^2\xi
\cos\theta_1 d\vp_1+ \f{1}{2}\sin^2\xi \cos\theta_2 d\vp_2, \ea and
\ba
ds^2_{CP^3}&=&d\xi^2+\cos\xi^2\sin^2\xi\left(d\psi+\f{\cos\theta_1}{2}d\vp_1-
\f{\cos\theta_2}{2}d\vp_2\right)^2 \no &&
+\f{1}{4}\cos^2\xi\left(d\theta_1^2+\sin^2\theta_1
d\vp_1^2\right)+\f{1}{4}\sin^2\xi(d\theta_2^2+\sin^2\theta_2
d\vp_2^2). \label{cp} \ea This expression (\ref{cp}) of $CP^3$ can
be found in e.g.  \cite{CLP}.

By comparing the above result with the conventional reduction
formula (below we always work with the string frame metric setting
$\al=1$) \be
ds^2_{11D}=e^{-2\phi/3}ds^2_{IIA}+e^{\f{4}{3}\phi}(d\ti{y}+\ti{A})^2,
\ee where $\ti{y}$ is compactified as $\ti{y}\sim \ti{y}+2\pi$.
Since we are taking the $Z_k$ orbifold, we identify $\ti{y}=ky$,
which leads to the value of dilaton \be e^{2\phi}=\f{R^3}{k^3}=
2^{\f{5}{2}}\pi \s{\f{N}{k^5}}. \ee The RR 2-form $F^{(2)}=d\ti{A}$
in the type IIA string is explicitly given by \ba F^{(2)}&=&
k\Bigl(-\cos\xi\sin\xi d\xi \we
(2d\psi+\cos\theta_1d\vp_1-\cos\theta_2 d\vp_2)\no &&
-\f{1}{2}\cos^2\xi\sin\theta_1 d\theta_1\we d\vp_1
-\f{1}{2}\sin^2\xi\sin\theta_2 d\theta_2 \we d\vp_2\Bigr), \ea while
the RR 4-form remains the same \be F^{(4)}=\f{3R^3}{8}\ep_{AdS_4}.
\ee The string frame metric now becomes \be
ds_{IIA}^2=\ti{R}^2(ds_{AdS4}^2+4ds_{CP3}^2), \ee where
$\ti{R}^2=\f{R^3}{4k}=\pi\s{\f{2N}{k}}$. In this way we obtain the
$AdS_4\times CP^3$ IIA background \cite{ABJM,NPW}. This background
preserves the 24 supersymmetries including the near horizon
enhancement as it is dual to three dimensional ${\cal N}=6$
superconformal symmetry.

\subsection{Holographic Entanglement Entropy}

To measure the degrees of freedom in a given conformal field theory,
a useful quantity is known as the entanglement entropy $S_A$ in
addition to the ordinary thermodynamical entropy. We expect that it
becomes more important in $CFT_3$ since in odd dimensions we do not
have a precise definition of the central charges. We trace out the
subsystem $A$ which is defined by an infinite strip with the width
$l$. Then the holographic area law formula in \cite{RT} leads to the
following result\footnote{Here we employed the explicit value
Vol$(CP^3)=\f{\pi^2}{12}$ of the volume of $CP^3$ in the coordinate
(\ref{cp}).} from the analysis of minimal surfaces in the Poincare
$AdS_4$ \be S_A=\f{\s{2}}{6\pi}N^2\s{\f{k}{N}}\left(\f{L}{a}
-2\pi\f{\Gamma(3/4)^2}{\Gamma(1/4)^2}\cdot\f{L}{l}\right), \ee where
$L$ represents the infinitely large length of the strip and $a$
denotes the ultraviolet cutoff (or the lattice spacing). Since $S_A$
is proportional to $\f{1}{\s{\lambda}}$ in addition to the leading
factor $N^2$ in the planar limit, we cannot explain this result from
the free field theory approximation. Therefore we can say that this
system is a more interacting theory than the ${\cal N}=4$
Yang-Mills, where we can qualitatively reproduce the supergravity
result of $S_A$ from the free Yang-Mills \cite{RT}.

\section{Penrose Limit of Type IIA on $AdS_4\times CP^3$}
\setcounter{equation}{0} \subsection{Penrose Limit and Plane Wave
Solution} We would like to take the Penrose limit \cite{BMN} of type
IIA background $AdS_4\times CP^3$. We express the metric of $AdS_4$
by \be ds^2=-\cosh^2\rho dt^2+d\rho^2+ \sinh\rho^2 d\Omega_2^2. \ee
The metric of $CP^3$ is given by (\ref{cp}). We are focusing on the
null geodesic defined by \be \rho=0,\ \ \ \theta_1=\theta_2=0,\ \ \
\xi=\f{\pi}{4}.  \ee We introduce a new angular coordinate \be
\ti{\psi}=\psi+\f{\vp_1-\vp_2}{2}. \ee The Penrose limit is defined
by the following coordinate transformation \ba
\f{t+\ti{\psi}}{2}=x^+,\ \ \ \ti{R}^2\f{t-\ti{\psi}}{2}=x^{-},\ \ \
\rho=\f{r}{\ti{R}},\ \ \ \theta_i=\f{\s{2}y_i}{\ti{R}},\ \ \
\xi=\f{\pi}{4}+\f{y_3}{2\ti{R}}, \ea setting $\ti{R}$ to infinity
with $x^\pm,r,y_1,y_2$ and $y_3$ kept finite.

In the end we find the following metric in this limit
$\ti{R}\to\infty$ \ba ds^2_{IIA}&=&-4dx^+dx^-
-(r^2+y_3^2)(dx^+)^2+dx^+(-y_1^2d\vp_1+y_2^2d\vp_2)\no &&+
dr^2+r^2d\Omega_3^2+(dy_1^2+y_1^2d\vp_1^2)+(dy_2^2+y_2^2d\vp_2^2)
+dy_3^2. \ea At the same time, the RR fluxes becomes \be
F_{+y_3}=\f{k}{2\ti{R}},\ \ \ \ F_{+r\Omega_2}=\f{3k}{2\ti{R}}r^2.
\ee If we define \be \ti{\vp}_1=\vp_1 -\f{x^+}{2},\ \ \
\ti{\vp}_2=\vp_2 +\f{x^+}{2}, \label{redefa} \ee we can rewrite the
metric as \ba ds^2_{IIA}&=&-4dx^+dx^-
-\left(r^2+y_3^2+\f{y_1^2+y_2^2}{4}\right)(dx^+)^2 \no &&+
dr^2+r^2d\Omega_3^2+(dy_1^2+y_1^2d\ti{\vp_1}^2)+(dy_2^2+y_2^2d\ti{\vp}_2^2)
+dy_3^2, \label{plane} \ea which is a familiar form of the plane
wave.

If we introduce the Cartesian coordinate $(x_1,\ddd,x_8)$ in an
obvious way we get \be ds^2=-4dx^+dx^- -\left(\sum_{i=1}^4
x_i^2+\f{1}{4}\sum_{i=5}^8x_i^2\right)(dx^+)^2+\sum_{i=1}^8(dx^i)^2,
\ee with \be F_{+4}=\f{k}{2\ti{R}},\ \ \
F_{+123}=\f{3}{2}\f{k}{\ti{R}}. \label{ppw}\ee Since the dilaton is
expressed as $e^{\phi}=\f{2\ti{R}}{k}$, we can rewrite the values of
RR-fluxes as $e^{\phi}F_{+4}=1$ and $e^{\phi}F_{+123}=3$, which will
be useful later.

This plane-wave background (\ref{ppw}) in IIA string has been known
in the literature \cite{pp} and has been shown to have 24
supersymmetries as we expect.

\subsection{Gauge Theory Interpretation}

It is argued that the type IIA on $AdS_4\times CP^3$ is dual to the
't Hooft limit of ${\cal N}=6$ superconformal Chern-Simons theory
with the level $(k,-k)$ and the gauge group $U(N)\times U(N)$ in
\cite{ABJM}. Since the gauge theory coupling in Chern-Simons
theories is proportional to $\f{1}{k}$, the 't Hooft coupling is
identified with $\lambda=\f{N}{k}$. Thus the 't Hooft limit is
defined as the large $N$ limit with $\lambda=\f{N}{k}$ kept finite.
It is natural to expect that our Penrose limit should correspond to
a certain limit of this gauge theory.

The ABJM theory consists of the Chern-Simons  $U(N)\times U(N)$
gauge potentials at level $(k,-k)$ coupled to the four chiral
superfields $A_1,A_2,B_1$ and $B_2$, whose structure is very
similar\footnote{ The Penrose limit of the Klebanov-Witten theory
$AdS_5\times T^{1,1}$ has been studied in \cite{PPO}.} to the
Klebanov-Witten theory \cite{KW}. The fields
$(A_1,A_2,\bar{B}_1,\bar{B}_2)$ belong to the $({\bf{N}},{\bf
\bar{N}})$ representation under the $U(N)\times U(N)$ gauge group
and they transform as the fundamental representation under the
$SU(4)$ R-symmetry of this ${\cal N}=6$ Chern-Simons theory.

First we relate the transverse scalars in the directions of
$(X^1,X^2,X^3,X^4)$ in (\ref{angles}) with the scalar
fields\footnote{In this paper we also express the scalar field part
of the chiral superfield $A_i$ and $B_i$.}
$(A_1,A_2,\bar{B}_1,\bar{B}_2)$ in the ABJM theory, following the
$SU(4)$ R-symmetry. We denote the conformal dimension $\Delta$ and
define $U(1)$ parts of R-charges $J_1,J_2,J_3$ as follows (here we
still did not perform the shift $\vp_i\to \ti{\vp_i}$ in
(\ref{redefa})) \ba J_1&=&-i\f{\de}{\de
\vp_1}\Bigl|_{\ti{\psi}}=-i\left(\f{\de}{\de\vp_1}
-\f{1}{2}\f{\de}{\de\psi}\right), \no J_2&=&-i\f{\de}{\de
\vp_2}\Bigl|_{\ti{\psi}}=-i\left(\f{\de}{\de\vp_2}
+\f{1}{2}\f{\de}{\de\psi}\right),\no J_3&=&-i\f{\de}{\de \psi}. \ea
Notice that in the final forms of three R-charges we fixed
$\ti{\psi},\vp_1$ and $\vp_2$ to be constant. Using the dependence
of the angles in (\ref{angles}) we find \ba &&J_1(A_1)=\f{1}{4},\ \
\ J_1(A_2)=-\f{3}{4}, \ \ \ J_1(B_1)=-\f{1}{4}, \ \ \
J_1(B_2)=-\f{1}{4},\no && J_2(A_1)=\f{1}{4},\ \ \ J_2(A_2)=\f{1}{4},
\ \ \ \ \ J_2(B_1)=-\f{1}{4}, \ \ \ J_2(B_2)=\f{3}{4},\no
&&J_3(A_1)=\f{1}{2},\ \ \ J_3(A_2)=\f{1}{2}, \ \ \ \ \
J_3(B_1)=\f{1}{2}, \ \ \ \ \  J_3(B_2)=\f{1}{2}.  \ea

Now we would like to relate the light-cone momenta $p^+$ and $p^-$
in the type IIA string to the gauge theoretic quantities assuming
the $AdS_4/CFT_3$ duality. To do this we need to rewrite the metric
in terms of $\ti{\vp}_i$ instead of $\vp_i$ as in (\ref{plane}).  In
this process, we regard any derivative as the one with
$\ti{\psi},\ti{\vp}_1$ and $\ti{\vp}_2$ fixed to be a constant. In
the end, we find \be 2p^-=i\f{\de}{\de x^+}=\Delta-J, \ \ \ \
2p^+=i\f{\de}{\de x^-}=\f{\Delta+J}{\ti{R}^2}, \ee where $J$ is
defined by \be J=J_3+\f{1}{2}J_1-\f{1}{2}J_2. \ee Explicitly, we get
\be J(A_1)=\f{1}{2},\ \ \ J(A_2)=0, \ \ \ J(B_1)=\f{1}{2}, \ \ \
J(B_2)=0. \label{rch} \ee

\subsection{World-Sheet Analysis}

First we analyze the bosonic sector. The world-sheet action in the
light-cone gauge $X^+=2p^+\tau$ looks like (notice $0\leq \sigma
\leq \pi$) \ba S_{B}&=&\f{1}{4\pi\al}\int d\sigma d\tau \de_a X^\mu
\de_a X^\nu g_{\mu\nu}(X),\no &=& \f{1}{4\pi\al}\int d\sigma d\tau
\left[\sum_{i=1}^8\left((\de_\tau X^i)^2-(\de_\sigma
X^i)^2\right)-4(p^+)^2\sum_{i=1}^4 (X^i)^2 -(p^+)^2\sum_{i=5}^8
(X^i)^2\right]. \no \ea  Then we easily find that the spectrum is
given by (setting $\al=1$) \be 2p^-_{B}=\sum_{n=-\infty}^\infty
N^{(1)}_n \s{1+\f{n^2}{(p^+)^2}}+ \sum_{n=-\infty}^\infty N^{(2)}_n
\s{\f{1}{4}+\f{n^2}{(p^+)^2}}, \ee where $N^{(1)}$ (and $N^{(2)}$)
denote the total occupation number of $n$-th string modes with
respect to the oscillators $\ap^{1,2,3,4}_{n}$ (and
$\ap^{5,6,7,8}_{n}$). We always need to impose the level matching
condition \be \sum_{n= -\infty}^\infty n(N^{(1)}_n+N^{(2)}_n)=0. \ee

Using the relation $p^+=\f{\Delta+J}{2\ti{R}^2}\simeq
\f{J}{\ti{R}^2}$, we can rewrite the above formula in term of the
gauge theory quantities \be \Delta-J= \sum_{n=-\infty}^\infty
N^{(1)}_n \s{1+\f{2\pi^2 n^2}{J^2}\cdot\f{N}{k}}+
\sum_{n=-\infty}^\infty N^{(2)}_n \s{\f{1}{4}+\f{2\pi^2
n^2}{J^2}\cdot\f{N}{k}}. \label{spec} \ee

As in the BMN case \cite{BMN}, we expect that the insertion of the
string oscillators corresponds to that of the impurity operators in
Tr$(A_1B_1)^J$. Indeed, $A_1B_1$ and their powers are the unique
operators which satisfy\footnote{Here we neglect the contribution
from the operators with Wilson line attached \cite{ABJM} since we
are assuming $k$ is large.} $\Delta-J=0$, as is clear from
(\ref{rch}). Also notice that Tr$(A_1B_1)^J$ is the chiral primary
operator. By inspecting the $R$-charge of impurities we can easily
identify (assuming the zero mode $n=0$) the 4 oscillators
$\ap^{5,6,7,8}_0$ with \be A_1B_2,\ \ \ A_1\bar{A_2},\ \ \ A_2B_1,\
\ \ ,\bar{B_2}B_1. \label{osc} \ee Indeed these four operators
satisfy $\Delta-J=\f{1}{2}$. Therefore we argue that the oscillators
$(\ap^{5}_0-i\ap^{6}_0,\ap^{5}_0+i\ap^{6}_0,
\ap^{7}_0-i\ap^{8}_0,\ap^{7}_0+i\ap^{8}_0)$ are dual to the
replacement procedures \be A_1\to A_2, \ \ B_1\to \bar{A}_2,\ \
A_1\to \bar{B}_2,\ \ B_1\to B_2. \ee On the other hand, we expect
that the three oscillators $\ap^{1,2,3}_{0}$ should be dual to the
covariant derivative $D_{\mu}$, where $\mu=0,1,2$. We still need to
identify one more. There are six other operators which satisfy
$\Delta-J=1$: \be A_1\bar{A}_1,\ \ \ \bar{A}_2 \bar{B}_2,\ \ \
A_2\bar{A}_2,\ \ \ \bar{B}_1B_1,\ \ \ \bar{B}_2B_2,\ \ \
\bar{B}_2\bar{A}_2.\ee Among them only $A_1\bar{A}_1$ and
$\bar{B}_2B_2$ are independent from the double excitations of the
previous operations in (\ref{osc}). Therefore, $\ap^{4}_{0}$ is
expected to be dual to a linear combination of these operators.

Next we turn to the fermionic sector (we follow the convention in
\cite{Michelson:2002ps}). The fermion part in the light cone gauge
$\Gamma^+S=0$ in the Green-Schwartz formalism looks like \be
S_{F}=\f{1}{4\pi\al}\int d\tau d\sigma \de_a X^\mu
\bar{S}\Gamma_\mu(\delta^{ab}-\ep^{ab}\Gamma^{11})D_bS, \ee where
$S$ is a ten dimensional Majorana spinor and the covariant
derivative $D_b$ is the pullback to the world-sheet of the
supercovariant derivative in IIA supergravity \be
D_\mu\ep=\nabla_\mu\ep+\f{e^{\phi}}{4}F_{\mu\nu}\Gamma^\nu\Gamma^{11}\ep
-\f{e^\phi}{(4!)^2}(3F_{\ap\beta\gamma\delta}\Gamma^{\ap\beta\gamma\delta}\Gamma_\mu
-F_{\ap\beta\gamma\delta}\Gamma_\mu\Gamma^{\ap\beta\gamma\delta})\ep.
\ee Then the action is simplified up to a constant \be S_{F}=\int
d\tau d\sigma\left[\bar{S}\Gamma_+(\de_\tau+\Gamma^{11}\de_\sigma)S
+\f{p^+}{2}\bar{S}\Gamma_+
(\Gamma^4\Gamma^{11}+3\Gamma^{123})S\right]. \ee The equation of
motion becomes \be (\de_\tau+\Gamma^{11}\de_\sigma)S=-\f{p^+}{2}
(\Gamma^4\Gamma^{11}+3\Gamma^{123})S. \ee By multiplying
$\de_\tau-\Gamma^{11}\de_\sigma$ we obtain \be
(\de_\tau^2-\de_\sigma^2)S=(2p^+)^2 \mu^2 S, \ee where \be
\mu^2\equiv
-\left(\f{\Gamma^4\Gamma^{11}+3\Gamma^{123}}{4}\right)^2. \ee

Since $16\mu^2=10-6\Gamma^{12341\!1}$, we can conclude that among
eight physical fermions, half of them have $\mu^2=1$, while other
half do $\mu^2=\f{1}{4}$. This mass spectrum is exactly the same as
the bosonic one. Thus we find the fermion spectrum \be
2p^-_{F}=\sum_{n=-\infty}^\infty N^{(1)}_n \s{1+\f{n^2}{(p^+)^2}}+
\sum_{n=-\infty}^\infty N^{(2)}_n \s{\f{1}{4}+\f{n^2}{(p^+)^2}}, \ee
and the values of $\Delta-J$ for fermions are given by the same
formula (\ref{spec}). Among totally sixteen fermions in the dual
gauge theory, four fermions satisfy $\Delta-J=\f{1}{2}$ and other
four fermions do $\Delta-J=\f{3}{2}$, while the rest eight fermions
have $\Delta-J=1$. Therefore we find that the string spectrum is
consistent with this gauge theory fermionic operators at least for
the zero modes.

In this way have shown a nice matching between the zero modes of the
IIA string theory in the Penrose limit and the gauge theory
operators. This is of course expected since the operators dual to
zero modes (or KK modes) are protected under the change of the
coupling constant.

\section{BMN like Operators}
\setcounter{equation}{0} Motivated by the analysis of Penrose limit
in the previous section we would like to examine non-BPS operators
in the ABJM theory. Especially, we are interested in the BMN-like
operators\footnote{The existence of the spin chain like structure
was already suggested in \cite{GaYi,ABJM}.} (almost BPS operators).
Indeed it is not difficult to find analogous operators (refer also
to \cite{GaYi} for similar operators in less supersymmetric
Chern-Simons theory.).

We would like to concentrate on the following operators assuming $J$
is very large \be {{\cal O}_n}=\f{1}{\s{2J}}\sum_{l=0}^J e^{2\pi
i\f{ln}{J}}\mbox{Tr}[(A_1B_1)^{l}A_1B_2 (A_1B_1)^{J-l}(A_1B_2)]. \ee
Notice that ${\cal O}_0$ is chiral primary since the index $i$ and
$j$ of $A_i$ and $B_j$ are both symmetrized independently. We can
treat the impurity of the form $A_2B_1$ exactly in the same way.

\subsection{Anomalous Dimension}

 We would like to compute two point functions of these operators
 and obtain the anomalous
dimensions to leading order. Since we know that the operator ${\cal
O}_0$ is chiral primary and its anomalous dimension is vanishing, we
have only to consider the Feynman diagrams whose results depend on
$n$. Then the relevant part of the Lagrangian looks like (we follow
the convention in \cite{ABJM}) \be {\cal
L}=\sum_{i=1}^2\left(\de_\mu A^i \de^\mu \bar{A}^i +\de_\mu B^i
\de^\mu \bar{B}^i\right)
+\f{16\pi^2}{k^2}\mbox{Tr}[B_2A_1B_1\bar{B}_2\bar{A}_1\bar{B_1}]
+\f{16\pi^2}{k^2}\mbox{Tr}[B_1A_1B_2\bar{B}_1\bar{A}_1\bar{B_2}].\label{lag}
\ee Then the propagator is normalized as follows \be \la
A^i_{a\bar{b}}(x)\bar{A}^j_{\bar{c}d}(0)\lb=\la
B^i_{\bar{b}a}(x)\bar{B}^j_{d\bar{c}}(0)\lb=\f{\delta_{ij}\delta_{d
a}\delta_{\bar{c}\bar{b}}}{4\pi|x|},\ee where we neglect any
interactions which do not affect our leading computation of the
anomalous dimension.

Since the two interactions in (\ref{lag}) exchange $B_2$ with the
two nearest $B_1$s, respectively, they produce the phase factors
$e^{\pm 2\pi i\f{n}{J}}$, which is very similar to the BMN analysis
\cite{BMN}. Also notice that the insertion of either of these
interactions adds two loops in the fat diagram and leads to $N^2$
factor. Therefore we obtain \be \la  {{\cal O}_n}(x){\overline{{\cal
O}}_n}(0)\lb =\f{{\cal N}}{|x|^{2(J+2)}}
\left(1+\f{1}{(4\pi)^3}\cdot \f{32\pi^2 N^2}{k^2}(\cos\f{2\pi
n}{J}-1)\cdot I(x)\right), \ee where $I(x)=|x|^3\int
\f{dy^3}{|y|^3|x-y|^3}$ and ${\cal N}$ is a normalization factor.
Since $I(x)\sim 8\pi \log x\Lambda$ ($\Lambda$ is the cutoff), we
can conclude that the leading anomalous dimension $\delta^{CS}_n$ of
${\cal O}_n$ is given by \be
\delta^{CS}_n=4\pi^2\f{N^2n^2}{k^2J^2}+\ddd, \label{ano} \ee where
$\ddd$ denotes the higher order terms with respect to $\f{N}{kJ}$.

It is also useful to remember that in the original analysis of BMN
operators there are two different coupling constants: one is the
rescaled 't Hooft coupling $\f{\lambda}{J^2}$ and the other is the
effective string coupling $\f{J^2}{N}$ \cite{Sca}. The latter
appears when we consider the non-planar diagrams, which we neglected
in the above. In our Chern-Simons gauge theory, we can see that the
non-planar corrections come with the same factor $\f{J^2}{N}$. Since
in our argument which derives (\ref{ano}), we keep the rescaled 't
Hooft coupling $\f{N}{kJ}$ a small value, the non-planar correction
is negligible if $N\ll k^2$.

\subsection{Comparison with IIA Plane Wave}

One may naively guess that the Penrose limit of the IIA string
studied in section 3 corresponds to the BMN-like limit assumed in
the previous subsection \be \f{N}{kJ}=\mbox{finite},\ \ \ \ N \ll
k^2, \ee as the analogous relation was true in the celebrated
duality between $AdS_5\times S^5$ and the four dimensional $N=4$
Yang-Mills theory \cite{BMN}. However, this does not seem to be the
case here, even though about the chiral primary operators there is a
nice matching between them as we have seen in the previous section.
In fact, the anomalous dimension found in the Penrose limit reads
(see (\ref{spec})) \be \delta^{IIA}_n= \f{2\pi^2Nn^2}{kJ^2}+\ddd,
\label{anoii} \ee for the impurities of $A_1B_2$ and $A_2B_1$. This
is different from the result (\ref{ano}) obtained from the IIA
string spectrum on the plane wave by the factor $\f{2N}{k}$. In this
string theoretic calculation in the Penrose limit, we need to keep
the string coupling $e^{2\phi}$ small and $p^+$ finite, which
requires \be \f{1}{(p^+)^2}\sim \f{N}{kJ^2}=\mbox{finite},\ \ \
e^{2\phi}\sim \s{\f{N}{k^5}}\ll 1. \label{peny} \ee

We would like to argue that the disagreement between the leading
anomalous dimensions (\ref{ano}) and (\ref{anoii}) is not a
contradiction but is due to the violation of the BMN scaling (a
similar phenomenon in other type IIA backgrounds has been pointed
out in \cite{LiMa}). Notice that the violation of BMN scaling in
this sort of computations (i.e. near BPS states to the leading order
of the large $J$ limit) does not occur in the $AdS_5/CFT_4$ duality
of the ${\cal N}=4$ super Yang-Mills theory.

In other words, we expect that the anomalous dimension of ${{\cal
O}_n}$ in the large $J$ limit of the ABJM theory is given by \be
\delta_n=f(\lambda)\f{n^2}{J^2}+\ddd, \ee in terms of a certain
function $f(\lambda)$ of $\lambda=\f{N}{k}$. Our results (\ref{ano})
and (\ref{anoii}) predict the following behaviors\footnote{It is not
difficult to find functions with these properties. Indeed, we can
consider functions like $f(\lambda)=4\pi^2\f{\lambda^2}{1+2\lambda}$
or $f(\lambda)=4\pi^2\f{\lambda^2}{\s{1+4\lambda^2}}$, for example.}
\be f(\lambda)\to 2\pi^2\lambda\ \ \ (\lambda\to \infty),\ \ \ \
\mbox{and} \ \ \ \ f(\lambda)\to 4\pi^2\lambda^2 \ \ \ (\lambda\to
0). \ee  It will be very interesting to compute the function
$f(\lambda)$ exactly from the Chern-Simons theory.

\section{Free ${\cal N}=6$ Chern-Simons Theory on $S^1\times S^2$}
\setcounter{equation}{0}
Obviously, another limit which we can take to make a given theory
simpler and more tractable is the weak coupling limit. We would like
to finish this paper by studying the weak coupling limit $k\to
\infty$ of the  ${\cal N}=6$ Chern-Simons theory on $S^1\times S^2$.
We will show that the Hagedorn/deconfinement transition will occur
in almost the same way as in the ${\cal N}=4$ free Yang-Mills on $S^1\times
S^3$ \cite{Sud,AMMPR}.

The original ABJM action in this limit becomes
\begin{align}\label{ABJMaction}
  S_{ABJM} &= \int d^3x \f{1}{4\pi}\mbox{Tr}\left[ \left( A_{(1)}\we dA_{(1)} +
      \f{2}{3\s k}A_{(1)}^3\right) - \left(A_{(2)}\we dA_{(2)} +
      \f{2}{3\s k}A_{(2)}^3\right)\right]\no
  &+ \mbox{Tr} \sum_{i=1,2}\left[ |D_\mu^{(+)} A_i|^2 + |D_\mu^{(-)} B_i|^2 + i\bar\psi_i
    \Slash D^{(+)}\psi_i + i\bar\chi_i\Slash D^{(-)}\chi_i\right]
  + {\cal O} \left(\f{1}{k}\right),
\end{align}
where we define the covariant derivatives
\begin{align}
  D_\mu^{(\pm)} = \nabla_\mu \pm \f{i}{\s k}(A_{(1)\,\mu}\otimes
  \bold{1}- \bold{1}\otimes A_{(2)\,\mu}).
\end{align}
We must be careful in taking $k$ infinity since a naive treatment
spoils the Gauss' law constraint \cite{AMMPR}. We must choose the
gauge fixing by the temporal gauge
\begin{align}\label{gaugefix}
  A_{(a)\,0}(x) = \s k \,a_{(a)}, \quad (a=1,2).
\end{align}
Under this gauge fixing, the action (\ref{ABJMaction}) on $S^1\times
S^2$ becomes
\begin{align}
  S_{free} & = 
  iS_{CS}(A_{(1)};S^1\times S^2) - iS_{CS}(A_{(2)};S^1\times S^2)\no
  &+ \mbox{Tr} \sum_{i=1,2}\Big[ \bar{A}_i \left(\!\!
  -(D_\mu'^{(+)})^2\!\!
  +\!\!
    \f{\CR}{8} \right)A_i\! +\! \bar{B}_i \left(\!\! -(D_\mu'^{(-)})^2
    +\!\!
    \f{\CR}{8} \right)B_i \! +\! i\bar\psi_i
    \Slash D'^{(+)}\psi_i\! +\! i\bar\chi_i\Slash D'^{(-)}\chi_i\Big],
\end{align}
where we included $\f{\CR}{8}$ term which arises from a conformal
coupling of the scalar field and defined $D_\mu'^{(\pm)} =
(D_0'^{(\pm)}\equiv\de_0 \pm i(a_{(1)}\otimes
  \bold{1} - \bold{1}\otimes a_{(2)}), \nabla_1, \nabla_2)$.
The Ricci scalar is $\CR =2$ for the unit two sphere, and
integrating the matter fields out gives the one-loop effective
action
\begin{align}
  & \mbox{Tr}\ln\left(-(D_0'^{(\pm)})^2 - \nabla^2 + \f{\CR}{8}\right)
  = -\sum_{n=1}^{\infty}\f{1}{n}z_B(x^n)\, (\mbox{tr} U^n\,\mbox{tr} V^{-n}+\mbox{tr} U^{-n}\,\mbox{tr} V^{n}),\no
  & \mbox{Tr}\ln(-(\Slash D'^{(\pm)})^2)\! =\! Tr\ln(-(D_0'^{(\pm)})^2\!\! - \!\!\nabla^2\!\! +\!\! \f{\CR}{8})
  = \sum_{n=1}^{\infty}\f{(-)^{n+1}}{n}z_F(x^n)\,(\mbox{tr} U^n\,\mbox{tr} V^{-n}+\mbox{tr} U^{-n}\,\mbox{tr} V^{n}) ,
\end{align}
where we denote $x=e^{-\beta}$ and introduce
$U=e^{i\beta\alpha_{(1)}},\, V=e^{i\beta\alpha_{(2)}}$ as Wilson
loops along $S^1$. We omit the irrelevant terms independent of
$\alpha$ and define the single-particle partition function of bosons
and fermions as \be\label{SPF}
  z_B(x) = \f{x^{\f{1}{2}}(1+x)}{(1-x)^2}, \ \ \ \ z_F(x) = \f{2x}{(1-x)^2}.
\ee After all, the partition function becomes the expectation value
of the Wilson loops of $U(N)\times U(N)$ Chern-Simons gauge theory
\begin{align}
  Z =&\int [DA_{(1)}][DA_{(2)}]\exp \Big[ iS_{CS}(A_{(1)};S^1\times
    S^2) - iS_{CS}(A_{(2)};S^1\times S^2) \no
    &\qquad\qquad + \sum_{n=1}^\infty\f{1}{n}(4z_B(x^n)+(-)^{n+1}4z_F(x^n))(\mbox{tr} U^n\,\mbox{tr}
    V^{-n}+\mbox{tr} U^{-n}\,\mbox{tr} V^{n})\Big],\no
    =& \left\langle  \exp \left[
      \sum_{n=1}^\infty\f{1}{n}z_n(x)(\mbox{tr} U^n\,\mbox{tr}
    V^{-n}+\mbox{tr} U^{-n}\,\mbox{tr} V^{n})\right] \right\rangle_{S^1 \times S^2},
\end{align}
where, in the second equality, we define
$z_n(x)=4z_B(x^n)+(-)^{n+1}4z_F(x^n)$. It is known that only singlet
representation of the Wilson loops takes non-zero expectation value
in Chern-Simons gauge theory on $S^1\times S^2$ \cite{WiCS}, then we
can rewrite the above expression as the matrix model
\begin{align}
  Z = \int [dU][dV]\exp \left[
      \sum_{n=1}^\infty\f{1}{n}z_n(x)(\mbox{tr} U^n\,\mbox{tr}
    V^{-n}+\mbox{tr} U^{-n}\,\mbox{tr} V^{n})\right].
\end{align}
Once taking the large-$N$ limit, we can obtain the effective action
\begin{align}\label{MMaction}
  I_{eff} &= N^2 \sum_{n=1}^\infty \f{1}{n}\left(|u_n|^2 + |v_n|^2 -
  z_n(x)(u_nv_{-n}+u_{-n}v_n)\right),
\end{align}
where $u_n\equiv \mbox{tr} U^n/N,\, v_n\equiv \mbox{tr} V^n/N$ and
the first two terms in the right hand side come from the measure.

We now consider the saddle point of the matrix model action
(\ref{MMaction}). The eigenvalues $\lambda$ of the quadratic form in
(\ref{MMaction}) with respect to $(u_n,v_n)$, read $\lambda=1\pm
z_n(x)$. Thus the trivial saddle point $u_n=v_n=0$ is dominated if
$z_1(x)<1$ since $z_n(x)$ is monotonically decreasing function of
$n$. For $z_1(x)>1$, one of the eigenvalues becomes negative and the
action is dominated by another saddle point which gives order $N^2$
free energy. Then, there is a deconfinement transition at $z_1(x)=1$
and the Hagedorn temperature is calculated using (\ref{SPF}) as
  $T_H = \f{1}{\log (17+12\s 2)} \sim 0.283648$.

In this way we have shown that in the large $k$ limit (free limit),
a Hagedorn/deconfinement transition occurs in the ABJM theory. In
the strong coupling region $\f{N}{k}\gg 1$, this is expected from
supergravities \cite{Wi}: both the IIA string on $CP^3$ and the
M-theory on $S^7/Z_k$ have the $AdS_4$ black hole solution. To
understand the finite $k$ region in the gauge theory side, which is
dual to the M-theory, we need to take the non-singlet flux
contributions \cite{ABJM} into account and this will be an
interesting future problem.

\section{Conclusion}
\setcounter{equation}{0}
In this paper we examined the Penrose limit of the type IIA string
on $AdS_4\times CP^3$, which is argued to be dual to the ${\cal
N}=6$ Chern-Simons gauge theory (ABJM theory) in the 't Hooft limit.
We obtained the resulting plane wave background and compute the
string spectrum in terms of gauge theoretic quantities. For BPS
operators, we find the agreement between the IIA string and the ABJM
theory. Also the string spectrum in the plane wave limit provides us
with an important prediction of the anomalous dimensions in certain
sectors which satisfy $J\sim \s\f{N}{k}$ in the ABJM theory. We also
analyzed the gauge theory sides and argued that we can define
BMN-like (almost BPS) operators when the R-charge $J$ is large. We
calculated the leading anomalous dimensions for these BMN-like
operators and found that the results are different from the ones
computed in the IIA string on the plane wave. This shows that the
BMN scaling in the ABJM theory is violated already in this near BPS
sector as opposed to the ${\cal N}=4$ super Yang-Mills theory.
 This issue definitely deserves
future studies.

We also examined the weak coupling limit $k\to \infty$ of the ABJM
theory on $S^1\times S^2$ and evaluated the partition function at
finite temperature. We showed that the Hagedorn/deconfinement
transition occurs in this limit of the ABJM theory as naturally
expected.

\vskip3mm

\noindent {\bf Acknowledgments} TT would like to thank Y. Matsuo and
S. Terashima for useful conversations about the BLG theory. We are
also very grateful to H. Lin for helpful comments. The work of TN is
supported in part by JSPS Grant-in-Aid for Scientific Research
No.19$\cdot$3589.
 The work of TT is supported in part by JSPS Grant-in-Aid for
Scientific Research No.20740132, No.18840027 and by JSPS
Grant-in-Aid for Creative Scientific Research No. 19GS0219.

\vskip2mm



\begin{thebibliography}{99}

\baselineskip=10pt

\small


\bibitem{Sc}
  J.~H.~Schwarz,
  ``Superconformal Chern-Simons theories,''
  JHEP {\bf 0411}, 078 (2004)
  [arXiv:hep-th/0411077].

\bibitem{GaYi}
  D.~Gaiotto and X.~Yin,
  ``Notes on superconformal Chern-Simons-matter theories,''
  JHEP {\bf 0708}, 056 (2007)
  [arXiv:0704.3740 [hep-th]].

\bibitem{GaWi}
  D.~Gaiotto and E.~Witten,
   ``Janus Configurations, Chern-Simons Couplings, And The Theta-Angle in N=4
  Super Yang-Mills Theory,''
  arXiv:0804.2907 [hep-th].

\bibitem{BaLa}
  J.~Bagger and N.~Lambert,
  ``Modeling multiple M2's,''
  Phys.\ Rev.\  D {\bf 75}, 045020 (2007)
  [arXiv:hep-th/0611108];
``Gauge Symmetry and Supersymmetry of Multiple M2-Branes,''
  Phys.\ Rev.\  D {\bf 77}, 065008 (2008)
  [arXiv:0711.0955 [hep-th]];
  J.~Bagger and N.~Lambert,
  ``Comments On Multiple M2-branes,''
  JHEP {\bf 0802}, 105 (2008)
  [arXiv:0712.3738 [hep-th]].


\bibitem{Gu}
  A.~Gustavsson,
  ``Algebraic structures on parallel M2-branes,''
  arXiv:0709.1260 [hep-th];
``Selfdual strings and loop space Nahm equations,''
  JHEP {\bf 0804}, 083 (2008)
  [arXiv:0802.3456 [hep-th]].
















\bibitem{MuA}
  S.~Mukhi and C.~Papageorgakis,
  ``M2 to D2,''
  JHEP {\bf 0805}, 085 (2008)
  [arXiv:0803.3218 [hep-th]].




\bibitem{Bandres:2008vf}
  M.~A.~Bandres, A.~E.~Lipstein and J.~H.~Schwarz,
  ``N = 8 Superconformal Chern--Simons Theories,''
  JHEP {\bf 0805} (2008) 025
  [arXiv:0803.3242 [hep-th]].

\bibitem{Berman:2008be}
  D.~S.~Berman, L.~C.~Tadrowski and D.~C.~Thompson,
  ``Aspects of Multiple Membranes,''
  arXiv:0803.3611 [hep-th].


\bibitem{VanRaamsdonk:2008ft}
  M.~Van Raamsdonk,
  ``Comments on the Bagger-Lambert theory and multiple M2-branes,''
  JHEP {\bf 0805}, 105 (2008)
  [arXiv:0803.3803 [hep-th]].

\bibitem{LaTo}
  N.~Lambert and D.~Tong,
  ``Membranes on an Orbifold,''
  arXiv:0804.1114 [hep-th].

\bibitem{Morozov:2008cb}
  A.~Morozov,
  ``On the Problem of Multiple M2 Branes,''
  JHEP {\bf 0805} (2008) 076
  [arXiv:0804.0913 [hep-th]];
``From Simplified BLG Action to the First-Quantized
M-Theory,''
  arXiv:0805.1703 [hep-th].




\bibitem{Distler:2008mk}
  J.~Distler, S.~Mukhi, C.~Papageorgakis and M.~Van Raamsdonk,
  ``M2-branes on M-folds,''
  JHEP {\bf 0805}, 038 (2008)
  [arXiv:0804.1256 [hep-th]].

\bibitem{Gran:2008vi}
  U.~Gran, B.~E.~W.~Nilsson and C.~Petersson,
  ``On relating multiple M2 and D2-branes,''
  arXiv:0804.1784 [hep-th].

\bibitem{LiA}
  P.~M.~Ho, R.~C.~Hou and Y.~Matsuo,
  ``Lie 3-Algebra and Multiple M2-branes,''
  arXiv:0804.2110 [hep-th].

\bibitem{Gomis:2008cv}
  J.~Gomis, A.~J.~Salim and F.~Passerini,
  ``Matrix Theory of Type IIB Plane Wave from Membranes,''
  arXiv:0804.2186 [hep-th].

\bibitem{Bergshoeff:2008cz}
  E.~A.~Bergshoeff, M.~de Roo and O.~Hohm,
  ``Multiple M2-branes and the Embedding Tensor,''
  arXiv:0804.2201 [hep-th].

\bibitem{Hosomichi:2008qk}
  K.~Hosomichi, K.~M.~Lee and S.~Lee,
  ``Mass-Deformed Bagger-Lambert Theory and its BPS Objects,''
  arXiv:0804.2519 [hep-th].

\bibitem{LiB}
  G.~Papadopoulos,
  ``M2-branes, 3-Lie Algebras and Plucker relations,''
  JHEP {\bf 0805}, 054 (2008)
  [arXiv:0804.2662 [hep-th]].

\bibitem{LiC}
  J.~P.~Gauntlett and J.~B.~Gutowski,
  ``Constraining Maximally Supersymmetric Membrane Actions,''
  arXiv:0804.3078 [hep-th].


\bibitem{LiD}
  G.~Papadopoulos,
  ``On the structure of k-Lie algebras,''
  arXiv:0804.3567 [hep-th].


\bibitem{Ho:2008nn}
  P.~M.~Ho and Y.~Matsuo,
  ``M5 from M2,''
  arXiv:0804.3629 [hep-th].

\bibitem{DEA}
  J.~Gomis, G.~Milanesi and J.~G.~Russo,
  ``Bagger-Lambert Theory for General Lie Algebras,''
  arXiv:0805.1012 [hep-th].

\bibitem{DEB}
  S.~Benvenuti, D.~Rodriguez-Gomez, E.~Tonni and H.~Verlinde,
  ``N=8 superconformal gauge theories and M2 branes,''
  arXiv:0805.1087 [hep-th].


\bibitem{DEC}
  P.~M.~Ho, Y.~Imamura and Y.~Matsuo,
  ``M2 to D2 revisited,''
  arXiv:0805.1202 [hep-th].


\bibitem{Honma:2008un}
  Y.~Honma, S.~Iso, Y.~Sumitomo and S.~Zhang,
  ``Janus field theories from multiple M2 branes,''
  arXiv:0805.1895 [hep-th].

\bibitem{Fuji:2008yj}
  H.~Fuji, S.~Terashima and M.~Yamazaki,
  ``A New N=4 Membrane Action via Orbifold,''
  arXiv:0805.1997 [hep-th].


\bibitem{Krishnan:2008zm}
  C.~Krishnan and C.~Maccaferri,
  ``Membranes on Calibrations,''
  arXiv:0805.3125 [hep-th].

\bibitem{Song:2008bi}
  Y.~Song,
  ``Mass Deformation of the Multiple M2 Branes Theory,''
  arXiv:0805.3193 [hep-th].

\bibitem{Jeon:2008bx}
  I.~Jeon, J.~Kim, N.~Kim, S.~W.~Kim and J.~H.~Park,
  ``Classification of the BPS states in Bagger-Lambert Theory,''
  arXiv:0805.3236 [hep-th].

\bibitem{LW}
  M.~Li and T.~Wang,
  ``M2-branes Coupled to Antisymmetric Fluxes,''
  arXiv:0805.3427 [hep-th].

\bibitem{Hosomichi:2008jd}
  K.~Hosomichi, K.~M.~Lee, S.~Lee, S.~Lee and J.~Park,
   ``N=4 Superconformal Chern-Simons Theories with Hyper and Twisted Hyper
  Multiplets,''
  arXiv:0805.3662 [hep-th].



\bibitem{Banerjee:2008pd}
  S.~Banerjee and A.~Sen,
  ``Interpreting the M2-brane Action,''
  arXiv:0805.3930 [hep-th].

\bibitem{Lin:2008qp}
  H.~Lin,
  ``Kac-Moody Extensions of 3-Algebras and M2-branes,''
  arXiv:0805.4003 [hep-th].



\bibitem{DED}
  J.~Figueroa-O'Farrill, P.~de Medeiros and E.~Mendez-Escobar,
  ``Lorentzian Lie 3-algebras and their Bagger-Lambert moduli space,''
  arXiv:0805.4363 [hep-th];
 ``Metric Lie 3-algebras in Bagger-Lambert theory,''
  arXiv:0806.3242 [hep-th].


\bibitem{DEE}
 M.~A.~Bandres, A.~E.~Lipstein and J.~H.~Schwarz,
  ``Ghost-Free Superconformal Action for Multiple M2-Branes,''
  arXiv:0806.0054 [hep-th].




\bibitem{Passerini:2008qt}
  F.~Passerini,
  ``M2-Brane Superalgebra from Bagger-Lambert Theory,''
  arXiv:0806.0363 [hep-th].


\bibitem{PaSo}
  J.~H.~Park and C.~Sochichiu,
  ``Single M5 to multiple M2: taking off the square root of Nambu-Goto
  action,''
  arXiv:0806.0335 [hep-th].


\bibitem{DEF}
  J.~Gomis, D.~Rodriguez-Gomez, M.~Van Raamsdonk and H.~Verlinde,
  ``The Superconformal Gauge Theory on M2-Branes,''
  arXiv:0806.0738 [hep-th].


\bibitem{Ahn:2008ya}
  C.~Ahn,
  ``Holographic Supergravity Dual to Three Dimensional N=2 Gauge Theory,''
  arXiv:0806.1420 [hep-th].

\bibitem{Ezhuthachan:2008ch}
  B.~Ezhuthachan, S.~Mukhi and C.~Papageorgakis,
  ``D2 to D2,''
  arXiv:0806.1639 [hep-th].

\bibitem{DEG}
  S.~Cecotti and A.~Sen,
  ``Coulomb Branch of the Lorentzian Three Algebra Theory,''
  arXiv:0806.1990 [hep-th].


\bibitem{Mauri:2008ai}
  A.~Mauri and A.~C.~Petkou,
  ``An N=1 Superfield Action for M2 branes,''
  arXiv:0806.2270 [hep-th].

\bibitem{Bergshoeff:2008ix}
  E.~A.~Bergshoeff, M.~de Roo, O.~Hohm and D.~Roest,
  ``Multiple Membranes from Gauged Supergravity,''
  arXiv:0806.2584 [hep-th].


\bibitem{Blau}
  M.~Blau and M.~O'Loughlin,
  ``Multiple M2-Branes and Plane Waves,''
  arXiv:0806.3253 [hep-th].









\bibitem{ABJM}
  O.~Aharony, O.~Bergman, D.~L.~Jafferis and J.~Maldacena,
  ``N=6 superconformal Chern-Simons-matter theories, M2-branes and their
  gravity duals,''
  arXiv:0806.1218 [hep-th].

\bibitem{Kl}
  M.~Benna, I.~Klebanov, T.~Klose and M.~Smedback,
  ``Superconformal Chern-Simons Theories and $AdS_4/CFT_3$ Correspondence,''
  arXiv:0806.1519 [hep-th].

\bibitem{Mi}
J.~ Bhattacharya and S.~Minwalla, ''Superconformal Indices for
${\cal N}=6$ Chern Simons Theories,''  arXiv:0806.3251 [hep-th].


\bibitem{BMN}
  D.~E.~Berenstein, J.~M.~Maldacena and H.~S.~Nastase,
  ``Strings in flat space and pp waves from N = 4 super Yang Mills,''
  JHEP {\bf 0204} (2002) 013
  [arXiv:hep-th/0202021].

\bibitem{pp}
  K.~Sugiyama and K.~Yoshida,
  ``Type IIA string and matrix string on pp-wave,''
  Nucl.\ Phys.\  B {\bf 644} (2002) 128
  [arXiv:hep-th/0208029];
 S.~j.~Hyun and H.~j.~Shin,
  ``N = (4,4) type IIA string theory on pp-wave background,''
  JHEP {\bf 0210} (2002) 070
  [arXiv:hep-th/0208074].

\bibitem{LiMa}
  H.~Lin and J.~M.~Maldacena,
  ``Fivebranes from gauge theory,''
  Phys.\ Rev.\  D {\bf 74} (2006) 084014
  [arXiv:hep-th/0509235].






\bibitem{CLP}
  M.~Cvetic, H.~Lu and C.~N.~Pope,
  ``Consistent warped-space Kaluza-Klein reductions, half-maximal gauged
  supergravities and CP(n) constructions,''
  Nucl.\ Phys.\  B {\bf 597} (2001) 172
  [arXiv:hep-th/0007109].

\bibitem{NPW}
 B.~E.~W.~Nilsson and C.~N.~Pope,
  ``Hopf Fibration Of Eleven-Dimensional Supergravity,''
  Class.\ Quant.\ Grav.\  {\bf 1}, 499 (1984);
  S.~Watamura,
   ``Spontaneous Compactification And Cp(N): SU(3) X SU(2) X U(1),
   Sin**2-Theta-W, G(3) / G(2) And SU(3) Triplet Chiral Fermions In
  Phys.\ Lett.\  B {\bf 136}, 245 (1984).


\bibitem{RT}
S.~Ryu and T.~Takayanagi,
  ``Holographic derivation of entanglement entropy from AdS/CFT,''
  Phys.\ Rev.\ Lett.\  {\bf 96} (2006) 181602
  [arXiv:hep-th/0603001];
``Aspects of holographic entanglement entropy,''
  JHEP {\bf 0608} (2006) 045
  [arXiv:hep-th/0605073].


\bibitem{KW}
  I.~R.~Klebanov and E.~Witten,
  ``Superconformal field theory on threebranes at a Calabi-Yau  singularity,''
  Nucl.\ Phys.\  B {\bf 536}, 199 (1998)
  [arXiv:hep-th/9807080].


\bibitem{PPO}
  N.~Itzhaki, I.~R.~Klebanov and S.~Mukhi,
  ``PP wave limit and enhanced supersymmetry in gauge theories,''
  JHEP {\bf 0203} (2002) 048
  [arXiv:hep-th/0202153];
J.~Gomis and H.~Ooguri,
  ``Penrose limit of N = 1 gauge theories,''
  Nucl.\ Phys.\  B {\bf 635} (2002) 106
  [arXiv:hep-th/0202157];
 L.~A.~Pando Zayas and J.~Sonnenschein,
  ``On Penrose limits and gauge theories,''
  JHEP {\bf 0205} (2002) 010
  [arXiv:hep-th/0202186].

\bibitem{Michelson:2002ps}
  J.~Michelson,
  ``A pp-wave with 26 supercharges,''
  Class.\ Quant.\ Grav.\  {\bf 19} (2002) 5935
  [arXiv:hep-th/0206204].






\bibitem{Sca}
C.~Kristjansen, J.~Plefka, G.~W.~Semenoff and M.~Staudacher,
  ``A new double-scaling limit of N = 4 super Yang-Mills theory and PP-wave
  strings,''
  Nucl.\ Phys.\  B {\bf 643} (2002) 3
  [arXiv:hep-th/0205033];
D.~Berenstein and H.~Nastase,
  arXiv:hep-th/0205048;
 N.~R.~Constable, D.~Z.~Freedman,
  M.~Headrick, S.~Minwalla, L.~Motl, A.~Postnikov and W.~Skiba,
  ``PP-wave string interactions from perturbative Yang-Mills theory,''
  JHEP {\bf 0207} (2002) 017
  [arXiv:hep-th/0205089].

\bibitem{Sud}
  B.~Sundborg,
  ``The Hagedorn transition, deconfinement and N = 4 SYM theory,''
  Nucl.\ Phys.\  B {\bf 573}, 349 (2000)
  [arXiv:hep-th/9908001].


\bibitem{AMMPR}
  O.~Aharony, J.~Marsano, S.~Minwalla, K.~Papadodimas and M.~Van Raamsdonk,
  ``The Hagedorn / deconfinement phase transition in weakly coupled large N
  gauge theories,''
  Adv.\ Theor.\ Math.\ Phys.\  {\bf 8} (2004) 603
  [arXiv:hep-th/0310285].








\bibitem{WiCS}
  E.~Witten,
  ``Quantum field theory and the Jones polynomial,''
  Commun.\ Math.\ Phys.\  {\bf 121}, 351 (1989).


\bibitem{Wi}
  E.~Witten,
   ``Anti-de Sitter space, thermal phase transition, and confinement in  gauge
  theories,''
  Adv.\ Theor.\ Math.\ Phys.\  {\bf 2}, 505 (1998)
  [arXiv:hep-th/9803131].























\end{thebibliography}
\end{document}